\documentclass{elsart}
\usepackage{amssymb,graphicx,ulem}

\begin{document}

\begin{frontmatter}

\title{Using dimension reduction to improve outbreak predictability of multistrain diseases}

\author{Leah B.~Shaw$^1$, Lora Billings$^2$, and Ira B.~Schwartz$^1$}

\address{$^1$Naval Research Laboratory, Plasma Physics Division, Nonlinear Systems Dynamics Section, Code 6792, Washington, DC 20375}

\address{$^2$Department of Mathematical Sciences, Montclair State University, Upper Montclair, NJ 07043}

\begin{abstract}
Multistrain diseases have multiple distinct coexisting serotypes
(strains).  For some diseases, such as dengue fever, the serotypes
interact by antibody-dependent enhancement (ADE), in which infection
with a single serotype is asymptomatic, but contact with a second
serotype leads to higher viral load and greater infectivity. We
 present and analyze a dynamic compartmental model for
multiple serotypes exhibiting ADE. 
Using center manifold techniques, we show how the dynamics rapidly collapses to a lower dimensional system.  Using the constructed reduced
model, we can explain previously observed synchrony between certain
classes of
primary and secondary infectives \cite{Schwartz05}.  Additionally,
we show numerically that
the center manifold equations apply even to noisy systems.  Both
  deterministic and stochastic versions of the model enable
prediction of asymptomatic individuals that are difficult to track
during an epidemic.  We also show how this technique may be applicable
to other multistrain disease models, such as those with
cross-immunity.
\end{abstract}

\begin{keyword}
center manifold analysis, epidemic models, multistrain disease, dengue
\end{keyword}

\end{frontmatter}

\maketitle

\section{Introduction}

In the study of infectious disease, a problem of interest is the
coexistence of multiple strains.  Examples of persistent co-circulating multistrain diseases
include influenza \cite{AndreasenLL97}, malaria \cite{GuptaTAD94}, and
dengue fever \cite{FergusonDA99}. More recently, the avian flu
viruses, including  H5N1, were reported to coexist in several
genotypes until 2004 \cite{Webster04}. Other examples of viruses possessing
multiple strains may be seen in corona viruses, such as severe acute
respiratory syndrome (SARS) \cite{Zhi05}.  The presence of multiple strains
adds greater complication to models of disease dynamics due to an
  increasing number of stages through different infection-recovery combinatorics.  Reducing the
dimensionality of the model is desirable, both to obtain a simpler
model and to understand how the strains interact.

In this paper, we focus on dengue fever, which has four co-circulating
serotypes.  It is believed that following infection with and recovery
from one serotype, cross-reactive antibodies act to enhance the
infectiousness of a subsequent infection by another serotype \cite{Vaughn00}.  This phenomenon is termed
antibody-dependent enhancement (ADE).  Primary infections are less
severe, often asymptomatic, but secondary infections are associated
with more serious illness and greater risk for dengue hemorrhagic
fever (DHF).

Multistrain diseases with ADE-induced dynamics have been
modeled previously \cite{FergusonAG99,Schwartz05,Cummings05}.
Models of multistrain steady state endemic behavior and its stability
were considered for two strains in \cite{Feng97}, where both mosquito
population and humans were included. Although ADE was not explicitly
modeled, the authors did consider different parameter values of contact
susceptibility. Values of contact susceptibility less than unity model
cross-immunity, whereas values greater than unity model increased
susceptibility due to secondary infection.  Ferguson \textit{et al.}
modeled two serotypes of dengue and showed that ADE can lead to
oscillations and chaotic behavior \cite{FergusonAG99}.  Schwartz
\textit{et al.} developed a similar model with non-overlapping
compartments for greater ease of interpretation \cite{Schwartz05}.
The bifurcation structure was obtained for both the autonomous model
and a model with seasonal forcing.  Chaotic solutions were found for a
 wide range of parameter values.  It was observed from numerical simulations
that outbreaks of the four serotypes could occur asynchronously but
that certain primary and secondary infective compartments remained
synchronized. In particular, all compartments currently infected with
a particular strain exhibited in-phase outbreaks.  Cummings \textit{et
al.}, using the same model, explored the competitive advantage that
ADE confers to viruses.  Other models for dengue fever (e.g.,
\cite{EstevaV03}) have included the mosquito vector as in \cite{Feng97} but have 
not included the ADE effect.

Several models of multistrain diseases with cross-immunity rather than
ADE have appeared in the literature
\cite{AndreasenLL97,Gog02,EstevaV03}.  In these models, individuals
who have recovered from infection with one serotype have reduced
susceptibility to infection with other serotypes.  Oscillatory
dynamics have been observed.

Summarizing the results to date, in models of secondary infections where
  either ADE or cross-immunity is considered, the steady state
  equilibrium is not always stable, and persistent oscillations may
occur. For models with ADE, although ADE is the cause of oscillation onset, more complicated chaotic
or stochastic behavior is observed over a much wider range of
parameter values than those of periodic oscillations
\cite{Schwartz05}. Currently, there is no theory for predicting the
interrelationship between primary and secondary infective compartments.

In this paper, we consider the problem of revealing the dynamical
  structure between the secondary and primary infectives.
  Specifically, we analyze the model presented \cite{Schwartz05} and
  demonstrate that the dynamics can be reduced to a lower dimensional
  model.  Our analysis motivates the in-phase behavior previously
  reported for primary and secondary infectives.  The lower
  dimensional system allows the prediction of primary infective levels
  from knowledge of other compartment values.  This result may be
  useful in disease monitoring because little epidemiological data
  exists for the asymptomatic primary infectives.  We show numerically
  that our predictions hold even in noisy systems.  We also discuss
  the relevance of this technique to other multistrain diseases that
  do not display ADE, such as influenza, in which infection with one
  strain yields partial immunity to other strains.

\section{Description of general $n$-serotype model}
\label{sec:model}

We study the model of \cite{Schwartz05} for $n$ co-circulating serotypes. An individual can contract each serotype only once and is assumed to gain immunity to all serotypes after infection from two distinct serotypes. This compartmental model divides the population into disjoint sets and follows the size of these compartments as a percentage of the whole population over time. The variable definitions are as follows: \begin{center}
\begin{tabular}{cl}
$s$ & Susceptible to all serotypes; \\ 
$x_{i}$ & Primary infectious with serotype $i$ \\ 
$r_{i}$ & Primary recovered from serotype $i$ \\ 
$x_{ij}$ & Secondary infectious, currently infected with serotype $j$, \\ 
         & but previously had $i$ $(i\ne j)$. 
\end{tabular}
\end{center}
The model is a system of $n^2+n+1$ ordinary differential equations describing the rates of change of
the population within each compartment:
\begin{eqnarray}
\frac{ds}{dt} & = & \mu-\beta s\sum_{i=1}^{n}\left(x_{i}+\phi_i \sum_{j\ne i}x_{ji}\right)-\mu_{d}s\label{eq:model susceptibles}\\
\frac{dx_{i}}{dt} & = & \beta s\left(x_{i}+\phi_i \sum_{j\ne i}x_{ji}\right)-\sigma x_{i}-\mu_{d}x_{i}\label{eq:model primary}\\
\frac{dr_{i}}{dt} & = & \sigma x_{i}-\beta r_{i}\sum_{j\ne i}\left(x_{j}+\phi_j \sum_{k\ne j}x_{kj}\right)-\mu_{d}r_{i}\label{eq:model recovered}\\
\frac{dx_{ij}}{dt} & = & \beta r_{i}\left(x_{j}+\phi_j \sum_{k\ne j}x_{kj}\right)-\sigma x_{ij}-\mu_{d}x_{ij}.\label{eq:model secondary}
\end{eqnarray}
The parameters $\mu$, $\mu_{d}$, $\beta$, and $\sigma$ denote birth, death, contact, and recovery rates, respectively.  Antibody-dependent enhancement is governed by the parameters $\phi_i$.  Rates of infection due to primary infectives are of the form $\beta s x_i$, as in a standard SIR model, while infection rates due to secondary infectives are weighted by the appropriate $\phi$, taking the form $\beta \phi_j s x_{ij}$.  When $\phi=1$, there is no ADE, and both primary and secondary infectives are equally infectious.  When $\phi>1$, the ADE appears as an enhancement factor in the nonlinear terms involving secondary infectives.

Although the contact rate $\beta$ could be time dependent (e.g., due to seasonal fluctuations in the mosquito vector population), we assume it to be constant here for simplicity.  The fixed parameters throughout the paper are given by: $\mu=0.02$, $\mu_d=0$ or $\mu_d=0.02$, $\beta=400$, and $\sigma=100$, all with units of $years^{-1}$.  These values are consistent with estimates used previously in modeling dengue \cite{Cummings05,Billings06}.  We assume all serotypes have equal ADE factors:  $\phi_i=\phi$ for all $i$.  The ADE factor has not been measured from epidemiological data, so we test our results for various ADE values.

\begin{figure}[tbp]
\includegraphics[width=3.5in,keepaspectratio]{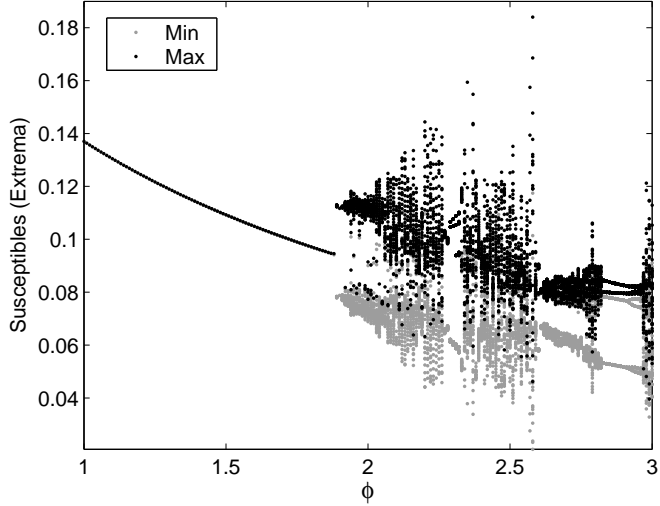}
\caption{ A bifurcation diagram for Eqs.~\ref{eq:model susceptibles}-\ref{eq:model secondary} with $n=4$, $\beta=400$, $\sigma=100$,  $\mu=0.02$, and $\mu_d=0.02$. Shown is a Poincar\'e section constructed from the extrema of a time series for susceptibles of length $t=100$ years after a transient of $t=5000$ years. The minima are denoted by grey points and the maxima are overlaid in black points. The Hopf bifurcation occurs at approximately $\phi=1.88333$.}
\label{fig:bifurcation}
\end{figure}

The dynamics of Eq.~\ref{eq:model susceptibles}-\ref{eq:model
  secondary} have been studied previously
\cite{Schwartz05,Billings06}.  A bifurcation diagram in the ADE factor
$\phi$ is given in Fig.~\ref{fig:bifurcation} for the four serotype
model.  The dynamics are qualitatively similar for other values of
$n$, the number of serotypes.  For $\phi \in [1,\phi_c)$, the system
  has a stable endemic steady state.  At a critical $\phi$ value, $\phi_c$, the system undergoes a Hopf bifurcation and begins to oscillate periodically.  For slightly larger $\phi$ values, the periodic solutions become unstable and the system oscillates chaotically.  Chaotic oscillations persist for most larger values for $\phi$, with the exception of narrow windows of stable periodic solutions.

It should be noted that previous studies of this model \cite{Schwartz05,Billings06} used a death rate equal to the birth rate ($\mu_d=\mu=0.02$) so that the population remained constant in time.  However, the endemic steady state cannot easily be obtained analytically when the mortality terms of Eqs.~\ref{eq:model susceptibles}-\ref{eq:model secondary} are included.  In our analysis, we omit the small mortality terms by setting $\mu_d=0$.  The dynamics with $\mu_d=0$ are qualitatively similar to that with $\mu_d=0.02$ and have the same bifurcation structure.  We demonstrate numerically that our results hold well even for the system with mortality.

The larger parameters $\beta,\sigma$ may be scaled by the small parameter $\mu$, defining
\begin{equation}
\beta=\beta_0/\mu,\sigma=\sigma_0/\mu,
\end{equation}
where $\beta_0,\sigma_0$ are $\mathcal{O}(1)$.  Eqs.~\ref{eq:model susceptibles}-\ref{eq:model secondary} with $\mu_d=0$ have an endemic steady state of
\begin{equation}
\left(s_0={\frac{\sigma_0}{\beta_0(\phi+1)}},x_{i,0}={\frac{{\mu}^{2}}{n \sigma_0}},
r_{i,0}={\frac{\sigma_0}{ \beta_0(n-1)(\phi+1)}},x_{ij,0}={\frac{{\mu}^{2}}{n(n-1)
\sigma_0}}\right)\label{eq:ss}\end{equation}
for all $i,j$.

\section{Center manifold analysis}

Although the full $n$-serotype model has $n^2+n+1$ dimensions, the attractor lies in a much lower dimensional space.  We
claim that the attractor actually lies in $2n+1$ dimensions to a good
approximation, due to a relationship between primary and secondary
infectives.  This result can be shown using center manifold analysis.
We now outline a derivation for the case of two serotypes and state the
center manifold equations for several other general cases of interest.

\subsection{$n=2$ serotypes}

We begin with Eqs.~\ref{eq:model susceptibles}-\ref{eq:model secondary} for two serotypes with no mortality ($\mu_d=0$).   We shift the variables so that the endemic fixed point (Eq.~\ref{eq:ss}) is at the origin:  $\bar{s}=s-s_0, \bar{x}_i=x_i-x_{i,0}, \bar{r}_i=r_i-r_{i,0}, \bar{x}_{ij}=x_{ij}-x_{ij,0}$.  The Jacobian in the shifted variables, evaluated at the origin, is given by 
\begin{equation}
\frac{1}{1+\phi}\left[\begin{array}{ccccccc}
\; 0 \; & -\sigma_0 & -\sigma_0 & \; 0 \; & \; 0 \; & \; -\phi \sigma_0 \; & \; -\phi\sigma_0 \; \\
0 & -\phi \sigma_0 & 0 & 0 & 0 & \phi\sigma_0 & 0 \\
0 & 0 & -\phi \sigma_0 & 0 & 0 & 0 & \phi \sigma_0 \\
0 & \sigma_0 (1+\phi) & -\sigma_0 & 0 & 0 & 0 & -\phi\sigma_0 \\
0 & -\sigma_0 & \; \sigma_0 (1+\phi) \; & 0 & 0 & -\phi\sigma_0 & 0 \\
0 & \sigma_0 & 0 & 0 & 0 & -\sigma_0 & 0 \\
0 & 0 & \sigma_0  & 0 & 0 & 0 & -\sigma_0
\end{array}\right]
\end{equation}
to zeroth order in $\mu$.

The Jacobian is not diagonalizable, as it has only 5 linearly independent eigenvectors.  However, a new set of variables $w$ is defined as follows:
\begin{eqnarray}
w= \frac{1}{1+\phi} \left[\bar{x}_{21}-\bar{x}_1, \bar{x}_{12}-\bar{x}_2,
(1+\phi) \bar{s}, \bar{x}_1+\phi \bar{x}_{21}, \bar{x}_2+\phi \bar{x}_{12}, \right. \nonumber \\
\left. \phi (\bar{x}_{21}-\bar{x}_1)-(1+\phi)\bar{r}_1,
 \phi (\bar{x}_{12}-\bar{x}_2)-(1+\phi)\bar{r}_2  \right] \label{changeofvars}
\end{eqnarray}
The transformation matrix for this change of variables contains the 5 linearly independent eigenvectors of the Jacobian and 2 additional vectors selected to be linearly independent and put the system in the desired form.

In the new variables, rescaling time to $\tau=t/\mu $, the dynamics are described by
\begin{eqnarray}
\frac{dw_1}{d\tau} &=&  -\sigma_0 w_1 +\beta_0 w_4 (w_7+\phi w_2 - w_3)+\frac{\mu^2 \beta_0}{2\sigma_0} (w_7+\phi w_2 - w_3) \label{dw1} \\
\frac{dw_2}{d\tau} &=& -\sigma_0 w_2 +\beta_0 w_5 (w_6+\phi w_1 -w_3) +\frac{\mu^2\beta_0}{2\sigma_0} (w_6+\phi w_1 -w_3) \label{dw2} \\
\frac{dw_3}{d\tau} &=& -\sigma_0 (w_4 +w_5) -\beta_0 (1+\phi) w_3 (w_4+w_5) -\frac{\mu^2 \beta_0}{\sigma_0}(1+\phi) w_3 \label{dw3} \\
\frac{dw_4}{d\tau} &=& \beta_0 w_4 (\phi^2 w_2+\phi w_7+w_3)+\frac{\mu^2 \beta_0}{2\sigma_0} (\phi^2 w_2+\phi w_7+w_3) \\
\frac{dw_5}{d\tau} &=& \beta_0 w_5 (\phi^2 w_1+\phi w_6+w_3)+\frac{\mu^2 \beta_0}{2\sigma_0} ( \phi^2 w_1+\phi w_6+w_3) \\
\frac{dw_6}{d\tau} &=& \sigma_0 (w_4-w_5) +\beta_0 w_4 (-\phi^2 w_2+\phi w_3 -\phi w_7 ) \nonumber \\
& & +\beta_0 w_5 \left[-\phi (1+\phi) w_1 -(1+\phi) w_6 \right] \nonumber \\
& & +\frac{\mu^2 \beta_0}{2\sigma_0} \left[-\phi^2 w_2+\phi w_3 -\phi w_7 -\phi (1+\phi) w_1 -(1+\phi) w_6\right] \\
\frac{dw_7}{d\tau} &=& \sigma_0 (w_5-w_4) +\beta_0 w_4 \left[-\phi(1+\phi) w_2 -(1+\phi) w_7 \right] \nonumber \\
& &+\beta_0 w_5 (-\phi^2 w_1 +\phi w_3 -\phi w_6) \nonumber \\
& & +\frac{\mu^2 \beta_0}{2\sigma_0} \left[ -\phi(1+\phi) w_2 -(1+\phi) w_7 -\phi^2 w_1 +\phi w_3 -\phi w_6 \right] \label{dw7} \\
\frac{d\mu}{d\tau} &=& 0 \label{dmu}
\end{eqnarray}
The Jacobian of Eqs.~\ref{dw1}-\ref{dw7}, evaluated at the origin and to zeroth order in $\mu$, is
\begin{equation}
\left[\begin{array}{ccccccc}
\; -\sigma_0 \; & 0 & 0 & \; 0 \; & \; 0 \; & \; 0 \; & \; 0 \; \\
0 & \; - \sigma_0 \; & 0 & 0 & 0 & 0 & 0 \\
0 & 0 & 0 & \; -\sigma_0 \; & \; -\sigma_0 \; & 0 & 0 \\
0 & 0 & 0 & 0 & 0 & 0 & 0 \\
0 & 0 & \; 0 \; & 0 & 0 & 0 & 0 \\
0 & 0 & 0 & \sigma_0 & -\sigma_0 & 0 & 0 \\
0 & 0 & 0  & -\sigma_0 & \sigma_0 & 0 & 0
\end{array}\right]
\end{equation}

Eqs.~\ref{dw1}-\ref{dmu} may be written in the compact form
\begin{eqnarray}
\dot{\mathbf{u}} &=&\mathbf{Au}+\mathbf{f}(\mathbf{u},\mathbf{v}, \mu) \nonumber \\
\dot{\mathbf{v}}&=&\mathbf{Bv}+\mathbf{g}(\mathbf{u},\mathbf{v}, \mu) \nonumber \\
\dot{\mu}&=&0 \nonumber
\end{eqnarray}
where $\mathbf{u}=(w_1,w_2),\mathbf{v}=(w_3,\ldots,w_7)$, where $\mathbf{A}$ and $\mathbf{B}$ are constant matrices such that the eigenvalues of $\mathbf{A}$ have negative real parts and the eigenvalues of $\mathbf{B}$ have zero real parts, and $\mathbf{f}$ and $\mathbf{g}$ are second order in $\mathbf{u},\mathbf{v},\mu$.  Therefore, center manifold theory \cite{CarrCM} can be applied.  The system will rapidly collapse onto a lower dimensional manifold defined by
\begin{eqnarray}
w_1 &=& h(\mu,w_3,w_4,\ldots,w_7) \\
w_2 &=& l(\mu,w_3,w_4,\ldots,w_7)
\end{eqnarray}

We expand $h$ and $l$ to second order
\begin{eqnarray}
w_1 &\approx& h_{0}+\sum_{i=3}^7 {h_{i} w_i}+h_\mu \mu + \sum_{i=3}^7 \sum_{j=i}^7 {h_{i,j} w_i w_j} +\sum_{i=3}^7 {h_{\mu,i} \mu w_i} \label{w1subs} \\
w_2 &\approx& l_{0}+\sum_{i=3}^7 {l_{i} w_i}+l_\mu \mu + \sum_{i=3}^7 \sum_{j=i}^7 {l_{i,j} w_i w_j} +\sum_{i=3}^7 {l_{\mu,i} \mu w_i}. \label{w2subs}
\end{eqnarray}
To solve for the coefficients of the expansion, we write equalities for $\frac{dw_1}{d\tau}$ and $\frac{dw_2}{d\tau}$.  This is done by equating the right hand sides of Eqs.~\ref{dw1}-\ref{dw2} with the time derivatives of Eqs.~\ref{w1subs}-\ref{w2subs}, using Eqs.~\ref{dw3}-\ref{dw7} to substitute for $\frac{dw_3}{d\tau},\ldots,\frac{dw_7}{d\tau}$ and Eqs.~\ref{w1subs}-\ref{w2subs} to substitute for $w_1,w_2$ as needed.  The coefficients of the expansion can be obtained by equating coefficients of like powers.

The coefficients $h_0,h_i,h_\mu,h_{\mu,i}$ and
$l_0,l_i,l_\mu,l_{\mu,i}$ are all 0, while some of the $h_{i,j}$ and
$l_{i,j}$ are nonzero.  After carrying out the above program, Eqs.~\ref{w1subs}-\ref{w2subs} simplify to
\begin{eqnarray}
w_1 & = & -\frac{\beta_0}{\sigma_0} w_3 w_4 -2\frac{\beta_0}{\sigma_0} w_4 w_5 + \frac{\beta_0}{\sigma_0} w_4 w_7 \label{w1} \\
w_2 & = &-\frac{\beta_0}{\sigma_0} w_3 w_5 -2\frac{\beta_0}{\sigma_0} w_4 w_5 + \frac{\beta_0}{\sigma_0} w_5 w_6 \label{w2}
\end{eqnarray}
when the coefficients are substituted.

Finally, we rewrite Eqs.~\ref{w1}-\ref{w2} in the original variables.  Thus we obtain the following approximation for the invariant manifold onto which the system collapses:
\begin{eqnarray}
\sigma_0 (\bar{x}_1 -\bar{x}_{21})=\beta_0 \left[\bar{s}-\bar{r}_2 +\frac{2-\phi}{1+\phi} \bar{x}_2+ \frac{3\phi}{1+\phi} \bar{x}_{12}  \right] (\bar{x}_1+\phi \bar{x}_{21}) \nonumber \\
\sigma_0 (\bar{x}_2 -\bar{x}_{12} )=\beta_0 \left[\bar{s}-\bar{r}_1  +\frac{2-\phi}{1+\phi} \bar{x}_1+ \frac{3\phi}{1+\phi} \bar{x}_{21} \right] (\bar{x}_2+\phi \bar{x}_{12}) \label{CM 2strain exact}
\end{eqnarray}
 The above equations for the center manifold are our main result
  of the paper. To bring out more of the structure, we make the
  following observations. We generally observed in numerical simulations that the infective compartments (and their deviations from the fixed point) are small compared to the susceptibles and recovereds.  Thus the infective correction terms to $\bar{s}-\bar{r}_i$ may be dropped to obtain a simpler expression for the center manifold:
\begin{eqnarray}
\sigma_0 (\bar{x}_1 -\bar{x}_{21})=\beta_0 \left(\bar{s}-\bar{r}_2 \right) (\bar{x}_1+\phi \bar{x}_{21}) \nonumber \\
\sigma_0 \left(\bar{x}_2 -\bar{x}_{12} \right)=\beta_0 (\bar{s}-\bar{r}_1) (\bar{x}_2+\phi \bar{x}_{12}) \label{CM 2strain approx}
\end{eqnarray}

\begin{figure}[tbp]
\includegraphics[width=4in,keepaspectratio]{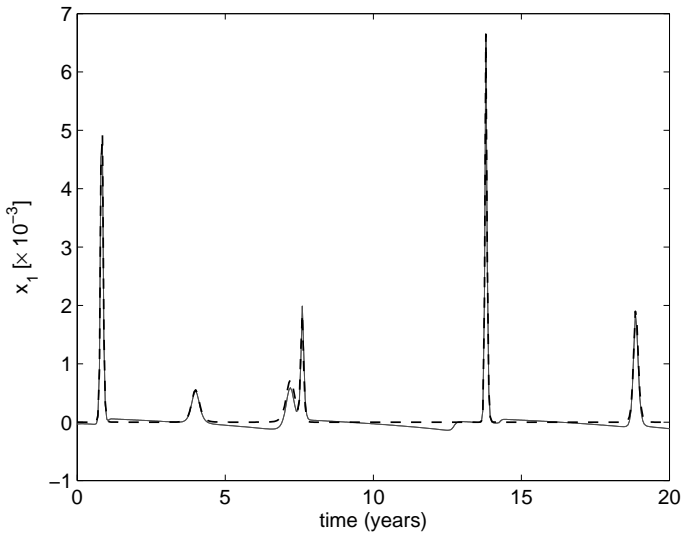}
\caption{Primary infectives with strain 1 vs.~time for two-strain model with no mortality.  Solid gray line:  prediction from approximate center manifold equations (predictions from Eqs.~\ref{CM 2strain exact} and \ref{CM 2strain approx} lie on same curve); dashed black line:  time trace from direct integration.  The curves from Eqs.~\ref{CM 2strain exact} and \ref{CM 2strain approx} coincide almost exactly.  The dynamics are chaotic with the parameter values used here:  $\beta=400$, $\sigma=100$, $\mu=0.02$, $\mu_d=0$, $\phi=2.8$.}
\label{fig:CM 2strain}
\end{figure}

Given values for the susceptibles, primary recovereds, and secondary
infectives, the primary infective compartments may be computed from
the approximate center manifold equations (either Eqs.~\ref{CM 2strain
  exact} or \ref{CM 2strain approx}).  Figure \ref{fig:CM 2strain}
compares the center manifold prediction for primary infectives with
that from simulations.  (Results from Eqs.~\ref{CM 2strain exact} and
\ref{CM 2strain approx} are nearly indistinguishable.)  The second
order approximation to the center manifold leads to reasonable
predictions for the primary infectives.  In particular, we accurately
predict the time and approximate magnitude of the bursts, which
correspond to epidemic outbreaks.  The prediction for the infectives
is occasionally negative because the center manifold equations are for
the $\bar{x}_i$, the deviations of infectives from their fixed point,
and sometimes the predicted deviations are large enough that adding
the fixed point $x_{i,0}$ yields negative values for the infective
variables $x_i$.  However, these errors are not physically important
because they occur in quiescent regions where epidemic outbreaks are
not occurring.  Moreover, the timing of the predicted outbreaks agrees
well with actual outbreak time of the full system.

\subsubsection{Asymmetric ADE factors}

Antibody enhancement  factors have not been measured experimentally, and it is possible that different serotypes could have different levels of enhancement.  We address the possibility of unequal ADE factors for the two serotype case.

\begin{figure}[tbp]
\includegraphics[width=4in,keepaspectratio]{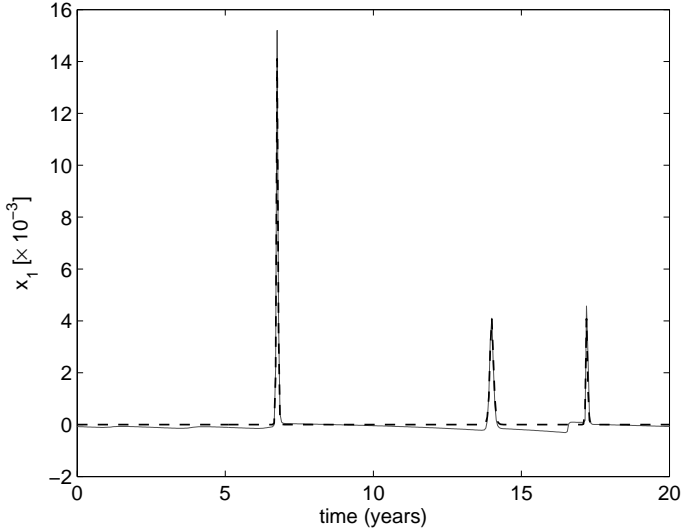}
\caption{Primary infectives with strain 1 vs.~time for two-strain model with no mortality and asymmetric ADE factors.  Solid gray line:  prediction from approximate center manifold equations (Eq.~\ref{asymADE1} and \ref{asymADE2}); dashed black line:  time trace from direct integration.  Parameters:  $\beta=400$, $\sigma=100$, $\mu=0.02$, $\mu_d=0$, $\phi_1=2.5$, $\phi_2=3.75$.  The ADE asymmetry measure is $\epsilon=0.5$.}
\label{fig:asymphi}
\end{figure}

A procedure similar to that used with symmetric ADE factors may be followed. In this case, an expansion for the location of the fixed point must also be used. For two strains with $\phi_1=\phi,\phi_2=\phi(1+\epsilon)$ and $\left|\epsilon\right| \ll 1$, the center manifold is approximated (to second order in the variables and $\mu,\epsilon$) by
\begin{eqnarray}
\sigma_0 \left[ \bar{x}_1-\bar{x}_{21}+{\epsilon \left( \bar{x}_1+\phi \bar{x}_{21} \right)  \frac {  \sigma_0}{
 2 \left( \sigma_0+1 \right)  \left( 1+\phi \right) }} \right] \;\;\;\;\;\; & \nonumber \\
\;\;\;\;\;\;  = \beta_0 \left[ \bar{s}-\bar{r_2}+{\frac { 2-
 \phi }{1+ \phi} \bar{x}_2}+{\frac {3\phi
}{1+ \phi }\bar{x}_{12}} \right]  \left( \bar{x}_1+\phi \bar{x}_{21} \right) \label{asymADE1} \\
\sigma_0 \left[ \bar{x}_2-\bar{x}_{12}-{ \epsilon
 \left( \bar{x}_2+\phi \bar{x}_{12} \right) \frac {\sigma_0 }
{ 2 \left( \sigma_0 +1 \right)  \left( 1+\phi \right) }} \right] \;\;\;\;\;\; & \nonumber \\
\;\;\;\;\;\;  =\beta_0 \left[ \bar{s}-\bar{r}_1+{\frac { 2-\phi }{1+\phi}\bar{x}_1}+{\frac {3\phi}{1+
\phi} \bar{x}_{21}} \right]  \left( \bar{x}_2+\phi \bar{x}_{12} \right) \label{asymADE2}
\end{eqnarray}
The center manifold equations again allow fairly accurate predictions of the primary infective compartments for small $\epsilon$ (data not shown).  Even when the $\left| \epsilon \right| \ll 1$ assumption does not hold, the time and amplitude of bursts in the primary infectives are generally predicted accurately (Fig.~\ref{fig:asymphi}).

\subsection{$n>2$ strains}

It is important to extend the analysis to a larger number of serotypes.  In particular, there are four observed serotypes for dengue fever.  The center manifold technique may again be applied for $n>2$ strains.  In the expressions for the center manifold, it is convenient to define the sum of secondary infectives currently infected with strain $k$:
\begin{equation}
\bar{z}_k = \sum_{i=1, i\neq k}^n {\bar{x}_{ik}} \label{zdef}
\end{equation}
For dengue fever, primary infectives are generally asymptomatic, and most hospital cases are secondary infectives.  The current infecting serotype can be determined from serology measurements \cite{NisalakENKTSBHIV03}.  $\bar{z}_k$, the sum of secondary infectives with serotype $k$, may be proportional to the number of hospital cases with serotype $k$.  This quantity may be relevant to disease monitoring in addition to providing a shorthand notation for expressing the center manifold equations.

We have completed center manifold analysis for $n=2,3,4$ serotypes.  The following equations for the center manifold summarize our results for $n=2,3,4$ and extrapolate them for general $n$:
\begin{eqnarray}
\sigma_0 \left[ \bar{x}_k -\bar{z}_k \right] &=& \beta_0 \left[ \bar{s} - \sum_{i\neq k} {\bar{r}_i} + f_k (\bar{x}_p,\bar{x}_{pq},n,\phi) \right] \left( \bar{x}_k + \phi \bar{z}_k \right)  \label{CM Nstrain x} \\
\hspace{-0.4in} \sigma_0 \left[ (n-1)\bar{x}_{jk} -\bar{z}_k \right] &=& \beta_0 \left[ (n-1)\bar{r}_j - \sum_{i\neq k} {\bar{r}_i} + g_{jk} (\bar{x}_p,\bar{x}_{pq},n,\phi) \right] \left( \bar{x}_k + \phi \bar{z}_k \right) \label{CM Nstrain z}
\end{eqnarray}
where
\begin{eqnarray}
f_k &=& \frac{n-(n-1)\phi}{(n-1)(1+\phi)} \left(\sum_{i\neq k} {\bar{x}_i}\right) + \frac{(2n-1)\phi}{(n-1)(1+\phi)} \left( \sum_{i,l, i\neq l } {\bar{x}_{il}} - \sum_{i\neq k} {\bar{x}_{ik}} \right) \\
g_{jk} &=& \frac{n-(n-1)\phi}{(n-1)(1+\phi)} \left( \sum_{i\neq k} {\bar{x}_i} -(n-1)\bar{x}_j \right) \nonumber \\
&& + \frac{(2n-1)\phi}{(n-1)(1+\phi)} \left( \sum_{ i,l, i\neq l } {\bar{x}_{il}} - \sum_{i\neq k} {\bar{x}_{ik}} -(n-1) \sum_{i\neq j} {\bar{x}_{ij}} \right)
\end{eqnarray}

We observe that the terms $f_k$ and $g_{jk}$ may be neglected.  They are linear functions of primary and secondary infective compartments.  At the fixed point, the infective compartments are order $\mu$, while the susceptibles and recovereds are order $1$ (Eq.~\ref{eq:ss}).  It therefore is reasonable that deviations of the infectives from the fixed point, represented by the variables $\bar{x}_p,\bar{x}_{pq}$, would generally be small compared to $\bar{s},\bar{r}_i$.  This assumption generally holds true in our simulations.

For each of the $n$ strains, Eqs.~\ref{CM Nstrain x} and \ref{CM Nstrain z} provide $n-1$ linearly independent equations, since some of the Eqs.~\ref{CM Nstrain z} are linearly dependent, allowing $n(n-1)$ dimensions to be eliminated.  Thus the dynamics of the system have dimension $2n+1$.  If the susceptible and primary recovered compartments are known, a single quantity $\bar{z}_k$ (Eq.~\ref{zdef}) for each strain is sufficient to describe the infective dynamics for that strain.  Primary and secondary infectives can then easily be computed using Eqs.~\ref{CM Nstrain x} and \ref{CM Nstrain z}.  We find that similar accuracy is attained when the $f_k,g_{jk}$ are set to $0$ for convenience as with the $f_k,g_{jk}$ terms retained.

\begin{figure}[tbp]
\includegraphics[width=4in,keepaspectratio]{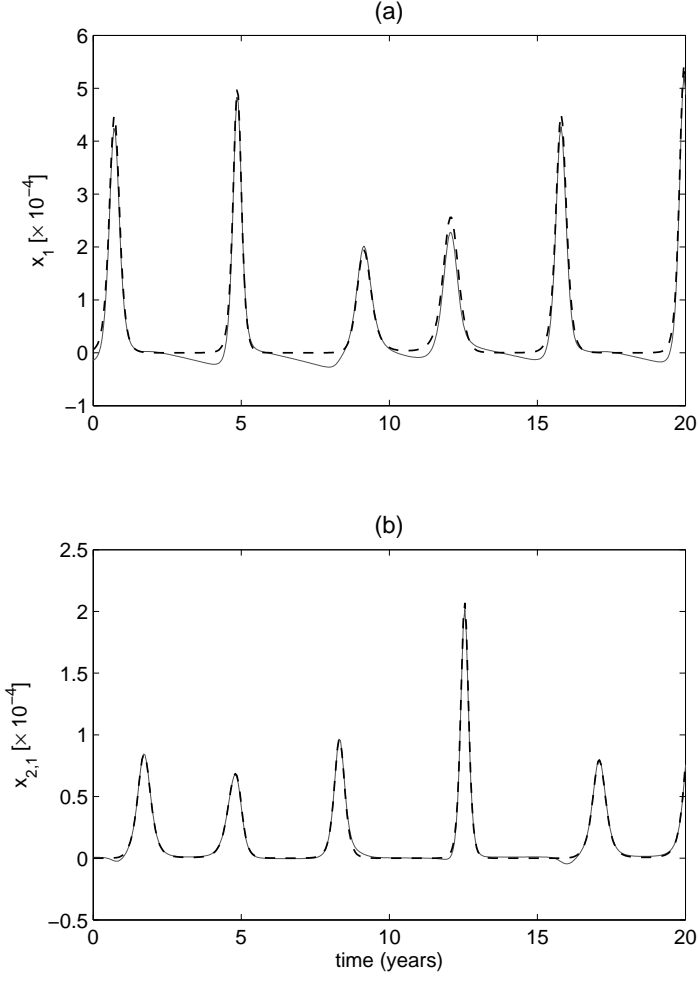}
\caption{(a) Primary infectives with strain 1 vs.~time for four-strain model with mortality; (b) secondary infectives contracting strain 1 then 2 vs.~time for four-strain model with mortality.  Solid gray lines:  prediction from approximate center manifold equations (Eqs.~\ref{CM Nstrain x} and \ref{CM Nstrain z} with $f_k,g_{jk}$ terms neglected); dashed black lines:  time traces from direct integration.  The dynamics are chaotic with the parameter values used here:  $\beta=400$, $\sigma=100$, $\mu=0.02$, $\mu_d=0.02$, $\phi=2.0$.}
\label{fig:CM 4strain}
\end{figure}

Figure \ref{fig:CM 4strain} compares the predictions from the center manifold equations (Eqs.~\ref{CM Nstrain x} and \ref{CM Nstrain z}) with actual values for sample primary and secondary infectives.  The full four-strain model with mortality ($\mu_d=0.02$) was used in the simulations.  Predictions were made by assuming that susceptibles, primary recovereds, and sums of secondary infectives ($\bar{z}_k$) were known.  Although the derivation of the center manifold equations omitted mortality terms, they generate accurate predictions for the full model.

\subsubsection{Phase synchrony between compartments}

It has been observed in numerical simulations that primary and secondary infectives currently infected with the same serotype oscillate in phase synchrony \cite{Schwartz05}.  Based on the center manifold reduction in dimension, we provide some analysis to motivate this effect.

We convert the shifted variables ($\bar{s},\bar{x}_i,\bar{r}_i,\bar{x}_{ij}$) to the original variables ($s,x_i,r_i,x_{ij}$) and use the center manifold equations \ref{CM Nstrain x}-\ref{CM Nstrain z} (with $f,g$ terms omitted as explained above) for substitution in the model ordinary differential equations, Eqs.~\ref{eq:model primary} and \ref{eq:model secondary}.  We obtain
\begin{eqnarray}
\frac{dx_{k}}{dt} & = &  -\sigma z_k +\beta \sum_{i\neq k } {r_i} \left( x_k +\phi z_k \right)  +\beta \left( s- \sum_{i\neq k } {r_i} \right) \left( x_{k,0} +\phi z_{k,0} \right) \label{dxk} \\
\hspace{-0.3in} (n-1) \frac{dx_{jk}}{dt} & = &  -\sigma z_k
 +\beta \sum_{i\neq k } {r_i} \left( x_k +\phi z_k \right) \nonumber \\
& & +\beta \left( (n-1) r_j - \sum_{i\neq k } {r_i} \right) \left( x_{k,0} +\phi z_{k,0} \right) \label{dxjk}
\end{eqnarray}
where we have defined
\begin{equation}
z_k = \sum_{i=1, i\neq k}^n {x_{ik}}
\end{equation}
for convenience.

Taking the difference between Eqs.~\ref{dxk} and \ref{dxjk} and converting back to rescaled variables yields
\begin{equation}
\frac{d\bar{x}_{k}}{dt}-(n-1) \frac{d\bar{x}_{jk}}{dt} = \beta \left[ \bar{s}-(n-1) \bar{r}_j \right] \left( x_{k,0} +\phi z_{k,0} \right) \label{derivdiff}
\end{equation}
The difference between Eqs.~\ref{CM Nstrain x} and \ref{CM Nstrain z} provides a useful substitution relation:
\begin{equation}
\sigma \left[ \bar{x}_k - (n-1) \bar{x}_{jk} \right] = \beta \left[ \bar{s} - (n-1) \bar{r}_j \right] \left( \bar{x}_k+\phi\bar{z}_{k}\right) \label{diffsubs} \label{CMdiffsubs}
\end{equation}

Substitution of Eq.~\ref{CMdiffsubs} into \ref{derivdiff} gives an equation for how the difference between primary and secondary infectives evolves in time:
\begin{equation}
\frac{d}{dt} \left[ \bar{x}_k(t) - (n-1) \bar{x}_{jk}(t) \right] = K \frac{ \bar{x}_k(t) - (n-1) \bar{x}_{jk}(t) }{  \bar{x}_k(t)+\phi\bar{z}_{k}(t) } \label{derivdiff2}
\end{equation}
where $K= \sigma ( x_{k,0} +\phi z_{k,0})$ is a positive constant of order $\mu$.  We restrict our consideration of Eq.~\ref{derivdiff2} to those intervals on which $\bar{x}_k(t)+\phi\bar{z}_{k}(t)$ is negative, thus avoiding singularities in $\left[ \bar{x}_k(t)+\phi\bar{z}_{k}(t)\right]^{-1}$.  That is, we consider intervals in which the sum of primary and secondary infectives with serotype $k$ (weighted by the ADE factor) is below the steady state value.  In the chaotic regime, these intervals occur during dropouts when serotype $k$ is present in low levels.  Such intervals cover the majority of the time domain and are interrupted by outbreaks of serotype $k$.  (Cf. Figure \ref{fig:CM 4strain}.)

Integrating Eq.~\ref{derivdiff2}, we find that
\begin{eqnarray}
\bar{x}_k(t) - (n-1) \bar{x}_{jk}(t) \hspace{2.9in} \nonumber \\
\;\;\;\;\;\;\;\; = \left[ \bar{x}_k(t_0) - (n-1) \bar{x}_{jk}(t_0) \right] \exp{ \left\{ K \int_{t_0}^t { \left[  \bar{x}_k(s)+\phi\bar{z}_{k}(s) \right]^{-1} ds } \right\} }
\end{eqnarray}
From our previous assumption, the argument of the integral is negative and is of order $1/\mu^2$, since the infectives range between 0 and the negative of the fixed point.  Therefore, $\left| \bar{x}_k(t) - (n-1) \bar{x}_{jk}(t)\right| \ll 1 $, and  $\bar{x}_k(t)$ and $ (n-1) \bar{x}_{jk}(t)$ approach each other rapidly, with their difference decreasing with $\exp{\left[-\mathcal{O}(1/\mu)\right]}$, where $\left|\mu\right|\ll 1$.

We have demonstrated that primary and secondary infectives with
serotype $k$ are phase synchronized, differing only by an
exponentially small term, during the intervals in which levels of
serotype $k$ are low.  Outbreaks in the infectives occur in bursts, so
although the relation $\left| \bar{x}_k(t) - (n-1)
\bar{x}_{jk}(t)\right| \ll 1 $ is not expected to hold during
outbreaks, the outbreaks have fast time scale dynamics. Therefore the duration of the outbreak is short enough that phase
synchrony is not lost.  It may be possible to examine the phase
locking during the outbreaks, but that would require a detailed
multiscale analysis, which is beyond the scope of the
present paper. Even so, our results help to explain the phase synchrony
previously observed between primary and secondary infectives currently
infected with the same strain \cite{Schwartz05}.

\section{ Stochastic perturbations}

We next consider Eqs.~\ref{eq:model susceptibles}-\ref{eq:model secondary} with multiplicative noise and discuss whether Eqs.~\ref{CM Nstrain x} and \ref{CM Nstrain z} can be applied to a noisy system.  We include the noise in our numerical integration as follows.

Let $\mathbf{Y}=\left\{s,x_i,r_i,x_{ij}\right\}_{i,j=1,\ldots,n}$ and $\dot{\mathbf{Y}}=\mathbf{F}(\mathbf{Y})$.  Define natural log coordinates $\mathbf{y}=\left\{y_i\right\}$, where $\ln(Y_i)=y_i$
and ${\bf \dot{y}}=\mathbf{f}({\bf y})$. Standard change of coordinates relates the two systems by:
\begin{equation}
 \dot{Y}_i =  Y_i f_i (\ln{\bf Y}), ~~ \mbox{since~~} \frac{\dot{Y}_i}{Y_i}=\dot{y}_i 
\end{equation}

We add noise to the natural log coordinates:
\begin{equation}
{\bf \dot{y}}=\mathbf{f}({\bf y})+\bf{\eta}. \label{eq:logvars}
\end{equation}
  Here, $\bf \eta$ is a vector of random variables with mean zero and standard deviation $\sigma_n$.  Transforming back to the original phase space, we obtain
\begin{eqnarray}
 \dot{Y}_i &=&  Y_i f_i (\ln{\bf Y})+\eta_i  Y_i \\
 \dot{Y}_i &=& F_i({\bf Y}) +\eta_i Y_i 
\end{eqnarray}
Therefore, the noise is multiplicative and scaled by each component.  Eq.~\ref{eq:logvars} was integrated numerically using a stochastic Euler method.

\begin{figure}[tbp]
\includegraphics[width=5in,keepaspectratio]{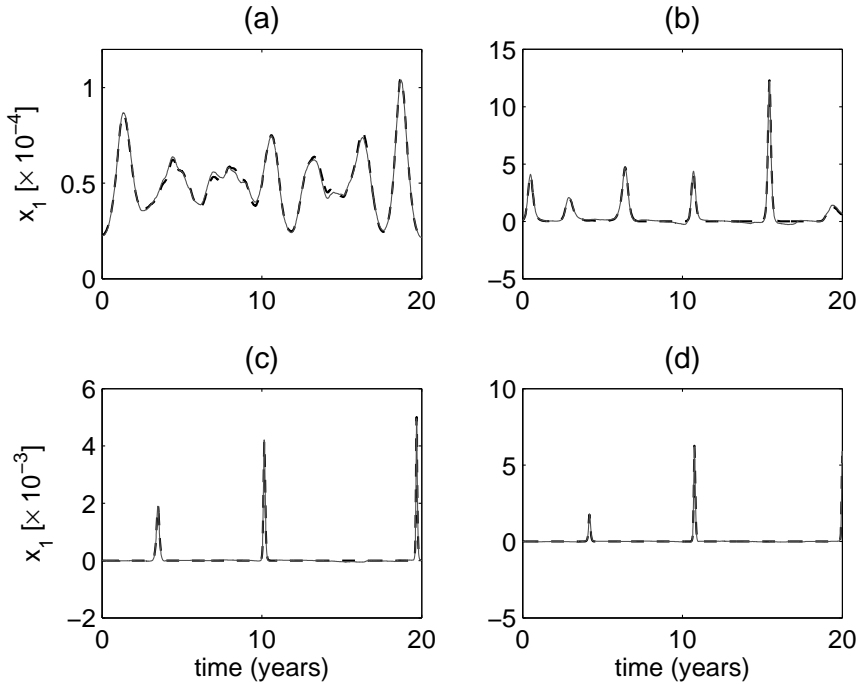}
\caption{
Primary infectives with strain 1 vs.~time for noisy four-strain model with no mortality, for various noise standard deviations.  Solid gray lines:  prediction from approximate center manifold equations (Eqs.~\ref{CM Nstrain x} and \ref{CM Nstrain z} with $f_k,g_{jk}$ terms neglected); dashed black lines:  time traces from direct integration. (a) $\sigma_n=0.01$, (b) $\sigma_n=0.05$, (c) $\sigma_n=0.1$, (d) $\sigma_n=0.15$.  Other parameter values:  $\beta=400$, $\sigma=100$, $\mu=0.02$, $\mu_d=0$, $\phi=1.9$.}
\label{fig:noisytt}
\end{figure}

The predictions from Eqs.~\ref{CM Nstrain x} and \ref{CM Nstrain z} hold approximately true for a wide range of noise standard deviations $\sigma_n$.  Sample results are shown in Figure \ref{fig:noisytt}.  For the parameter values in Figure \ref{fig:noisytt}, the endemic steady state is stable for $\sigma_n=0$.  Adding noise perturbs the system away from the steady state, and as the noise standard deviation $\sigma_n$ increases, the amplitude of the disease outbreaks increases.

\begin{figure}[tbp]
\includegraphics[width=3in,keepaspectratio]{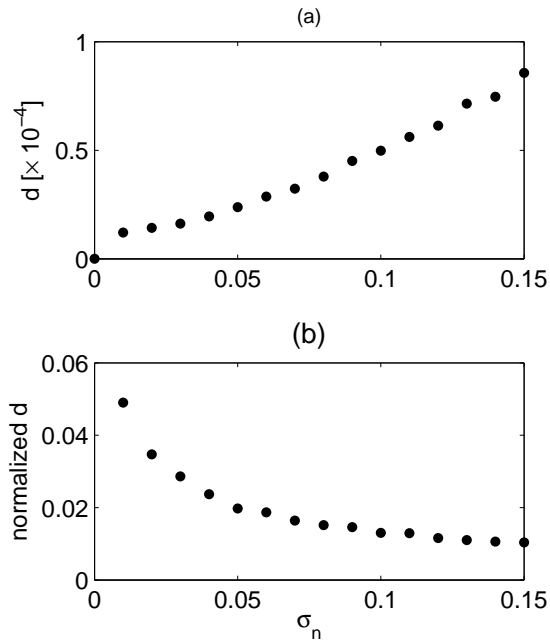}
\caption{(a) Average ``distance'' $d$ from center manifold vs. noise standard deviation; (b) normalized distance vs. noise standard deviation, where distance and normalized $d$ are as defined in the text.  Parameters:  $\beta=400$, $\sigma=100$, $\mu=0.02$, $\mu_d=0$, $\phi=1.9$.}
\label{fig:noisedist}
\end{figure}

To quantify the accuracy of the center manifold prediction for various $\sigma_n$, we define a metric for the ``distance'' of a trajectory from the center manifold:
\begin{equation}
d=\left< \sum_{i=1}^{n} { \left|x_i (t) - x_{i,pred} (t)  \right|  }\right>_t,
\end{equation}
where $x_{i,pred}$ is the predicted primary infective value using Eqs.~\ref{CM Nstrain x} and \ref{CM Nstrain z} (neglecting $f_k,g_{jk}$ terms) and $\left< \centerdot \right>_t$ denotes a time average.  We compute $d$ by sampling every 0.01 year for $10^4$ years (after first running for $10^4$ years to remove transients) in order to obtain good statistics.  Sample results are given in Figure \ref{fig:noisedist}(a).  The ``distance'' $d$ from the center manifold, i.e., the error in the center manifold prediction, increases as the noise increases.  However, the amplitude of the outbreaks also increases with larger noise, so it is not unexpected that $d$ would increase.  

We additionally compute a normalized distance in lieu of a percent error so that the effect of noise may be more carefully assessed.  (The actual percent error is not used because it would be greatly elevated by large percent errors in the dropout periods when the infectives are low.)  The $10^4$ year sampling period is divided into windows of 100 years, and the maximum value of the primary infective $x_1$ (arbitrarily chosen) is computed for each window.  These maxima give a scale for the outbreak amplitudes for each $\sigma_n$.  The distance $d$ is then divided by the average maximum for each $\sigma_n$ to compute a normalized distance from the center manifold.  Results are given in Figure \ref{fig:noisedist}(b).  Using this metric, we find that the normalized error in the center manifold prediction actually decreases in noisy systems.

\section{Conclusions and discussion}

We have analyzed a model for multistrain diseases with
antibody-dependent enhancement.  For a disease with $n$ strains and two sequential infections, the
model has $n^2+n+1$ equations.  Using center manifold analysis, we
have reduced the model to $2n+1$ dimensions.  Although we have derived an approximation to the manifold
on which the solutions for both deterministic and stochastic systems
lie, if we use this approximation and evolve the dynamics of a model
constrained to the manifold, we do not necessarily see the same
bifurcation structure. This is due to the fact that the analysis
itself is local and done near the steady state endemic point.  Despite that limitation, the lower dimensional system has other uses which we have described here.

We have partially explained the synchrony between primary and
secondary infective compartments currently infected with the same
serotype.  During the intervals when a serotype is present in low
levels, primary and secondary infectives of that serotype are
approximately related by a constant of proportionality.  Although our
derivation does not hold during outbreaks of that serotype, the
outbreaks occur in short bursts during which phase synchrony is not
lost.  As mentioned above, a full multiscale analytic treatment
may lend more insight into the phase relationship between primary and
secondary infections of a given serotype during the outbreak periods.

Our center manifold approximation enables prediction of each primary and secondary infective value if the disease parameters, susceptible and recovered compartments, and sums of secondary infectives (Eq.~\ref{zdef}) are known.  The prediction matches well with numerical simulations even when significant amounts of noise are added to the system.  It has been observed \cite{NisalakENKTSBHIV03} that most hospital cases of dengue fever are secondary infections.  From serotype measurements of these hospital cases, estimation of the number of secondary infectives that have a particular serotype $i$ should be possible.  On the other hand, there is little epidemiological data on the frequency of primary infections because patients with primary infections tend not to be as seriously ill and are not admitted to hospitals.  They are typically considered asymptomatic.  Obtaining primary infective data for dengue monitoring purposes might require random sampling of large numbers of apparently healthy individuals, which would not be feasible economically nor practically.  Our center manifold equations might be used to improve monitoring of dengue outbreaks by predicting compartments that cannot readily be measured, such as the primary infectives.

The fact that susceptible and recovered compartment data is necessary to make predictions for primary infectives is a potential drawback.  However, we observe in numerical simulations that the susceptibles and recovereds vary more slowly than the infective compartments, which quickly spike during an outbreak.  Therefore, it is possible that susceptible and recovered data could be obtained by sampling a population less frequently than would be required for infective data.  If such sampling is possible, the center manifold equations will yield information about primary infectives from indirect measurements from population data.

The applicability of our method to other models is also of interest.  Multistrain models for diseases with cross-immunity, in which recovery from one strain leads to reduced susceptibility to further infection with other strains, have been developed \cite{CastilloChavezHALL89,AndreasenLL97,Gog02,EstevaV03}.  We have analyzed a two-strain model with cross-immunity \cite{ShawUP}:
\begin{eqnarray}
\frac{ds}{dt} & = & \mu-\beta s\sum_{i=1}^{n}\left(x_{i}+\sum_{j\ne i}x_{ji}\right)-\mu_{d}s\label{crosssusceptibles}\\
\frac{dx_{i}}{dt} & = & \beta s\left(x_{i}+\sum_{j\ne i}x_{ji}\right)-\sigma x_{i}-\mu_{d}x_{i}\label{crossprimary}\\
\frac{dr_{i}}{dt} & = & \sigma x_{i}-\psi \beta  r_{i}\sum_{j\ne i}\left(x_{j}+\sum_{k\ne j}x_{kj}\right)-\mu_{d}r_{i}\label{crossrecovered}\\
\frac{dx_{ij}}{dt} & = & \psi \beta r_{i}\left(x_{j}+\sum_{k\ne j}x_{kj}\right)-\sigma x_{ij}-\mu_{d}x_{ij},\label{crosssecondary}
\end{eqnarray}
where $\psi<1$ expresses the degree of cross-immunity and $n=2$.  This two-strain cross-immunity model is introduced in \cite{CastilloChavezHALL89}.  This model differs from the ADE model in that rather than giving the secondary infectives a different infectivity (weighted by $\phi$), we give the primary recovereds a different susceptibility to further infection (weighted by $\psi$) than the susceptible compartment.  Using the coordinate transformation given in Eq.~\ref{changeofvars} (setting $\phi=1$), we can reduce Eqs.~\ref{crosssusceptibles}-\ref{crosssecondary} from 7 to 5 equations and show a relationship between primary and secondary infectives.  The eigenvalues of the Jacobian corresponding to rapidly collapsing directions are $n(n-1)$ eigenvalues equal to $-\sigma$, the recovery rate.  The center manifold analysis holds because the recovery rate is rapid compared to the birth rate.  It seems reasonable to expect that other models with cross-immunity might also be reduced and a relationship between primary, secondary, and perhaps higher order infectives might be derived.  Because cross-immunity models are often written in a different form with overlapping compartments (e.g.,\cite{AndreasenLL97,Gog02,EstevaV03}), some reformulation of the equations may be necessary before a center manifold approach can be used.

The results presented here were primarily for a symmetric multistrain model with equal disease parameters for each strain.  However, the case of unequal ADE factors was briefly considered, and a center manifold was found when the difference between ADE factors was a small parameter (Eqs.~\ref{asymADE1}-\ref{asymADE2}).  To model a multistrain disease with realistic parameters, it may be necessary to relax the symmetry constraint and derive results for an asymmetric system.  Such a derivation is unwieldy because the endemic steady state can no longer be determined analytically (except as an approximate expansion).  New techniques may be necessary to extend our current approach to more asymmetric models.

Another potentially interesting extension is to include spatial spread of the disease.  It has been observed from epidemiological data in Thailand that while dengue outbreaks within a single town may be of primarily one serotype, concurrent outbreaks in another town may have a different serotype predominant \cite{Endy02}.  A similar center manifold analysis might be attempted on a patch model, in which the population is divided into several spatially distinct patches that are coupled.

Research was supported by the Office of Naval Research
and the Center for Army Analysis. L.B. was supported
by the National Science Foundation under Grants
Nos. DMS-0414087 and CTS-0319555. L.B.S. is currently a
National Research Council postdoctoral fellow.

\bibliographystyle{plain}

\end{document}